# Theoretical evidence of H-He demixing under Jupiter and Saturn conditions

Xiaoju Chang[1,2*], Bo Chen[1,2*], Qiyu Zeng[1,2], Han Wang[3], Kaiguo Chen[1,2], Qunchao Tong[1,2], Xiaoxiang Yu[1,2], Dongdong Kang[1,2], Shen Zhang[1,2], Fangyu Guo[1,2], Yong Hou[1,2], Zengxiu Zhao[1,2], Yansun Yao[4**], Yanming Ma[5,6**] and Jiayu Dai[1,2 **]

[1]College of Science, National University of Defense Technology, Changsha 410073, China

[2]Hunan Key Laboratory of Extreme Matter and Applications, National University of Defense Technology, Changsha 410073, China

[3]Laboratory of Computational Physics, Institute of Applied Physics and Computational Mathematics, Beijing 100088, P. R. China

[4]Department of Physics and Engineering Physics, University of Saskatchewan, Saskatoon, Saskatchewan, Canada S7N 5E2

[5]State Key Lab of Superhard Materials and International Center for Computational Method and Software, College of Physics, Jilin University, Changchun 130012, China

[6]International Center of Future Science, Jilin University, Changchun 130012, China

* These authors contributed equally.

**Corresponding authors. E-mail addresses: yansun.yao@usask.ca, mym@jlu.edu.cn, jydai@nudt.edu.cn

## ABSTRACT

The immiscibility of hydrogen-helium mixture under the temperature and pressure conditions of planetary interiors is crucial for understanding the structures of gas giant planets (e.g., Jupiter and Saturn). While the experimental probe at such extreme conditions is challenging, theoretical simulation is heavily relied in an effort to unravel the mixing behavior of hydrogen and helium. Here we develop a method via a machine learning accelerated molecular dynamics simulation to quantify the physical separation of hydrogen and helium under the conditions of planetary interiors. The immiscibility line achieved with the developed method yields substantially higher demixing temperatures at pressure above 1.5 Mbar than earlier theoretical data, but matches better to the experimental estimate. Our results suggest a possibility that H-He demixing takes place in a large fraction of the interior radii of Jupiter and Saturn, *i.e.*, 27.5% in Jupiter and 48.3% in Saturn. This indication of an H-He immiscible layer hints at the formation of helium rain and offers a potential explanation for the decrease of helium in the atmospheres of Jupiter and Saturn.

The gravitational might of giant planets has played a key role in the formation of our solar system[1]. Jupiter and Saturn are the largest and most massive gas giants in the Sun's planetary system. Current models of Jupiter and Saturn suggest that the structures of both planets are similar in composition, both containing a visible cloud top, layers of gaseous hydrogen, liquid hydrogen, and metallic hydrogen, and possibly a rocky core[2,3]. Helium is present in all three layers of hydrogen, albeit with different abundances. The ratio of helium mass density to the sum of helium and hydrogen mass densities is $0.238\pm0.05$[4] in the atmosphere of Jupiter, and 0.06-0.08[5] in Saturn, both of which are conclusively lower than the estimated protosolar helium mass fraction ($0.275\pm0.01$[6]). The observed helium reduction is thought to be caused by the demixing of hydrogen and helium that precipitate toward deeper layers in the planet's interior. The sinking helium, through the exchange of gravitational potential energy to thermal energy, is thought to be an additional energy source to power Saturn's excess luminosity[7]. Thus, a complete diagram of the solubility of helium in hydrogen at planetary *P-T* conditions is highly required for an accurate modeling of Jupiter, Saturn, and other gas giants like them.

A recent experiment on giant planet modeling, through laser-driven shock compression of H-He mixtures, reveals a large region of H-He separation (~ 15% of the radial range) under Jovian interior conditions[8]. However, previous first-principles simulations[9–13] point toward much lower demixing temperatures and smaller immiscibility regions compared to the experiment. In particular, the demixing temperatures predicted by different theoretical models fall on either side of the adiabatic lines of Jupiter and Saturn with a large discrepancy of the order 2,000K. Such inconsistency would lead to completely different models for the planets' internal structures. In addition to the range of separation, the separation intensity is also important for planetary modeling - an *ab initio* calculation suggests excessive separation might cause Saturn's cooling time to be longer than the lifespan of the Solar System[14]. In theory, the miscibility range is usually determined based upon the Gibbs free energy of mixing ($\Delta G$), while different levels of theories (non-ideal entropy, choice of exchange-correlation functional) will yield slightly different results. On the other hand, the physical separation process plays an important role in the demixing of hydrogen and helium, but it has not been accounted in the previous studies. To this end, we will provide a new method to quantify the physical separation of H-He, and to determine the miscibility range directly from the separation process. We will also present a nonequilibrium approach for an improved evaluation of $\Delta G$ in large systems (~ 27,000 atoms), and use it to obtain the miscibility.

Previous density functional theory (DFT) evaluations of $\Delta G$ are usually carried out in systems with a small number of atoms (several tens to hundreds) to ensure sufficient mixing at all helium fractions and *P-T* conditions. This is because the conventional calculation method refers to equilibrium state, whereas a large system cannot represent the Gibbs free energy of a specific helium abundance once it undergoes demixing during equilibration. The difficulty of reaching equilibrium in *ab initio* molecular dynamics (MD) simulation for large systems should also be noted. Aside from low resolution, small systems tend to have non-negligible thermal fluctuations which may smear out the free energy difference between different configurations. One way to address this issue is to construct simulations with a sufficiently large number of atoms to enable statistical sampling, while at the same time maintaining a first-principles quality. Effective efforts including matter at extreme conditions were made by deep learning recently[15–17]. In this study, the MD simulation is substantially scaled up (to 27,000 atoms) using machine learning representation of potential energy surfaces. We constructed a deep-learning potential (DLP)[18] based on the strongly constrained and appropriately normed (SCAN) functional[19] with van der Waals (vdW) interaction from rVV10[20]. Details of the potential and its benchmarking are provided in Supplementary Information Note 2.

**Results**

The separation behaviors of H-He mixture are captured in the DLP accelerated MD simulations. However, a rigorous analysis of the MD trajectory must go beyond a visual observation to quantify the separation in the He-poor and He-rich zones. The intuitive mean-field methods, *i.e.*, calculating and averaging the thermodynamic properties in small subregions of the system, always yield unstable and non-convergent results. Since the subregions are treated as individuals, correlations among them are neglected and fluctuations are smoothed out. This is more problematic when dealing with extremely skewed compositions under real planetary conditions. Therefore, we designed an atomic miscibility analysis approach based on reweighted conditional probabilities ($P_{cond}$) for 'atoms in neighborhoods' to quantify the extent of separation in H-He mixtures. This method follows a similar statistical strategy to that of measuring the disorder in fluid mixture using conditional entropy[21]. Through tabulating the probability at which particular types of atoms are presented in the neighborhood of each atom in the systems, we can circumvent the aforementioned limitations and obtain a more precise numerical criterion for mixing and separation. We apply this method to the DFT-MD trajectories of 2,048 atoms with helium abundances of $X_{He}$ = 0.073 or 0.2. In addition to validating the accuracy of the DLP, it also demonstrates that the method can accurately differentiate helium abundance on a case-by-case basis, even when the separation levels are very close (see Fig .S11 in Supplementary Information Note 2). The validation of this method and its comparison to the mean field method are presented in Supplementary Information Note 3.

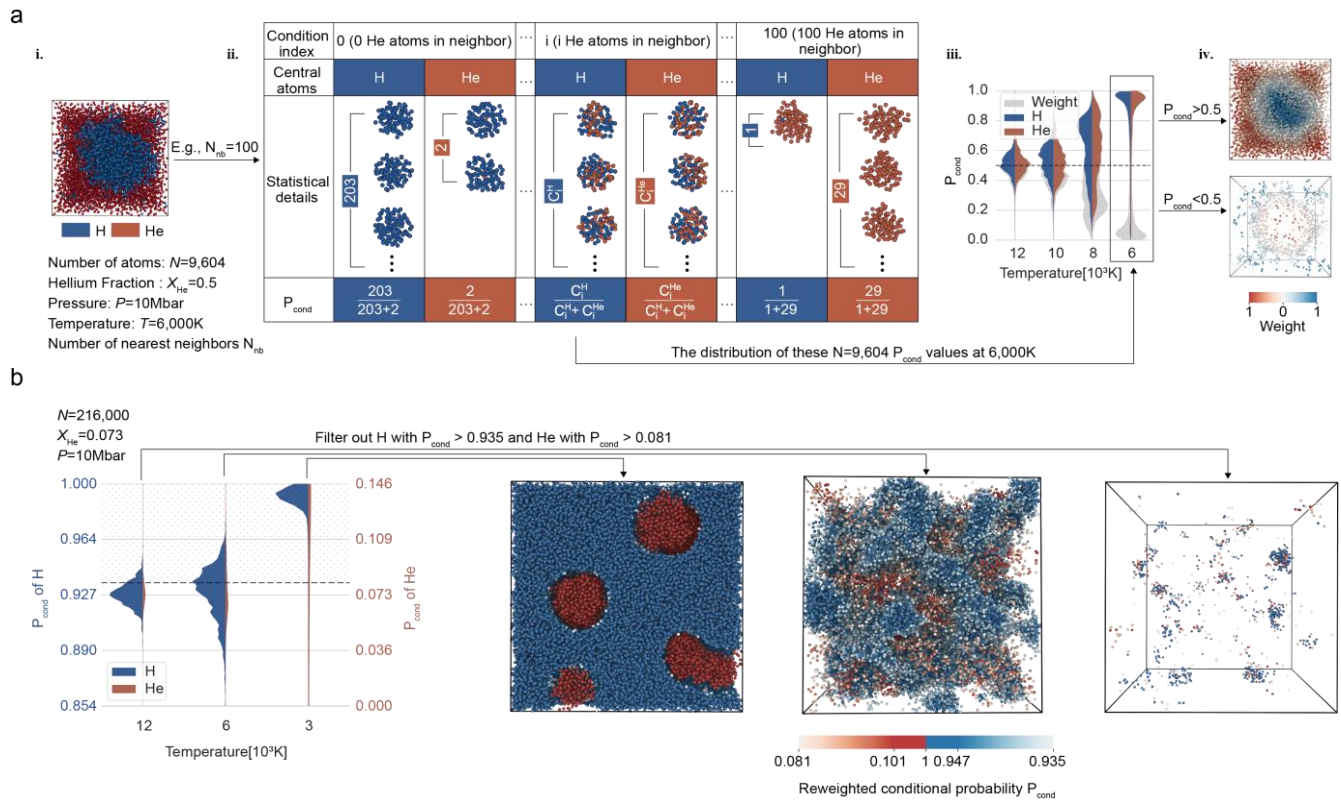

**Fig. 1. Schematic and utilization of Atomic Miscibility Analysis Based on Reweighted Conditional Probability (AMA-RCP). a,** Derivation of AMA-RCP method from molecular dynamics trajectories. From step i to ii, the neighborhood for a central atom is defined by the nearest $N_{nb}$ surrounding atoms. Here we take $N_{nb} = 100$ as an example. In step **ii**, all neighborhoods are classified based on their atomic compositions, spanning from all atoms being hydrogen to all atoms being helium. The $P_{cond}$ for Hydrogen/Helium in each class of neighborhood is the probability of a neighborhood in this class being centered by hydrogen/helium, for which the sum must be 1. Step **iii** demonstrates how the $P_{cond}$ distribution changes with temperature. For systems with $X_{He} = 0.5$, a $P_{cond}$ value greater or smaller than 0.5 indicates an atom is in a region enriched by itself or by the other element. The centrosymmetric gray shadows behind the distribution represent the weight of each atom corresponding to its $P_{cond}$ value when counting (See Methods for the explanation of re-weighting procedure). **iv** shows the reweighted classified results of the 6,000 K case. The color of each atom indicates its weight. **b,** An utilization example of AMA-RCP classification at an extremely skewed composition ($X_{He} = 0.073$). Three systems are prepared at the same

pressure at different temperatures. After obtaining $P_{cond}$ values of all atoms in each system, Helium atoms with $P_{cond}$ values greater than 0.081 and hydrogen atoms with $P_{cond}$ values greater than 0.935 are filtered out as a visualization of potential H-He demixing.

The main concept of AMA-RCP method is to determine the $P_{cond}$ for each atom in the system based on the atomic composition of the $N_{nb}$ nearest neighbors. All atoms are categorized by their $P_{cond}$ with a weight assigned for each category. With the categorized and reweighted results (Fig. 1a), $x_1$ and $x_2$ can be calculated directly,

$$\begin{cases} x_1 = \frac{N(\text{He}_{\text{He-poor}})}{N(\text{He}_{\text{He-poor}})+N(\text{H}_{\text{He-poor}})} \\ x_2 = \frac{N(\text{He}_{\text{He-rich}})}{N(\text{He}_{\text{He-rich}})+N(\text{H}_{\text{He-rich}})} \end{cases}. \quad (1)$$

The value $\Delta x = x_2 - x_1$ represents the abundance difference for helium in He-rich and He-poor zones, which is used as a measure of immiscibility (See Methods for the definition of immiscibility using $\Delta x$). Fig. 1b shows three representative examples with $X_{He} = 0.073$ for illustrating immiscible, weakly immiscible and miscible systems determined using this method. In the miscible system (12,000 K and 2 Mbar), the filtered hydrogen and helium atoms are few in number and gathered, indicating that phase separation does not occur, but rather the corresponding regions exhibit significant fluctuations. In contrast, the immiscible system filtered with the same $P_{cond}$ shows the prevalence of phase separation in the simulation box.

With the AMA-RCP, we conduct a rigorous analysis of mixing and separation behaviors in the MD trajectories of large H-He mixtures calculated using LAMMPS[22] package with DLP (See Methods for simulation details). Long-time trajectories obtained after the systems have reached equilibrium are used as inputs to obtain the degree of immiscibility for each system. Multiple helium abundances are used, including $X_{He} = 0.073$ and 0.089 calculated with 27,000-atom simulation boxes and $X_{He} = 0.2, 0.357, 0.5, 0.643, 0.8$, and 0.91 with 9,604-atom simulation boxes. The MD simulations are carried out at various temperatures and pressures to construct the miscibility diagram, for which a total of 529 MD trajectories are generated. To compare with previously published results[9,13], we also calculated the miscibility lines at 4 and 10 Mbar using thermodynamic integration for calculating free energy $\Delta G$ with 64-atom systems as the equation,

$$\Delta G(X_{He}) = G(X_{He}) - X_{He}G(1) - (1 - X_{He})G(0). \qquad (2)$$

The values of $x_1$ and $x_2$ can be determined by applying a common tangent construction to $\Delta G$. The calculated $\Delta G$ and corresponding comparison with results of the previous work[13] are presented in the Methods section.

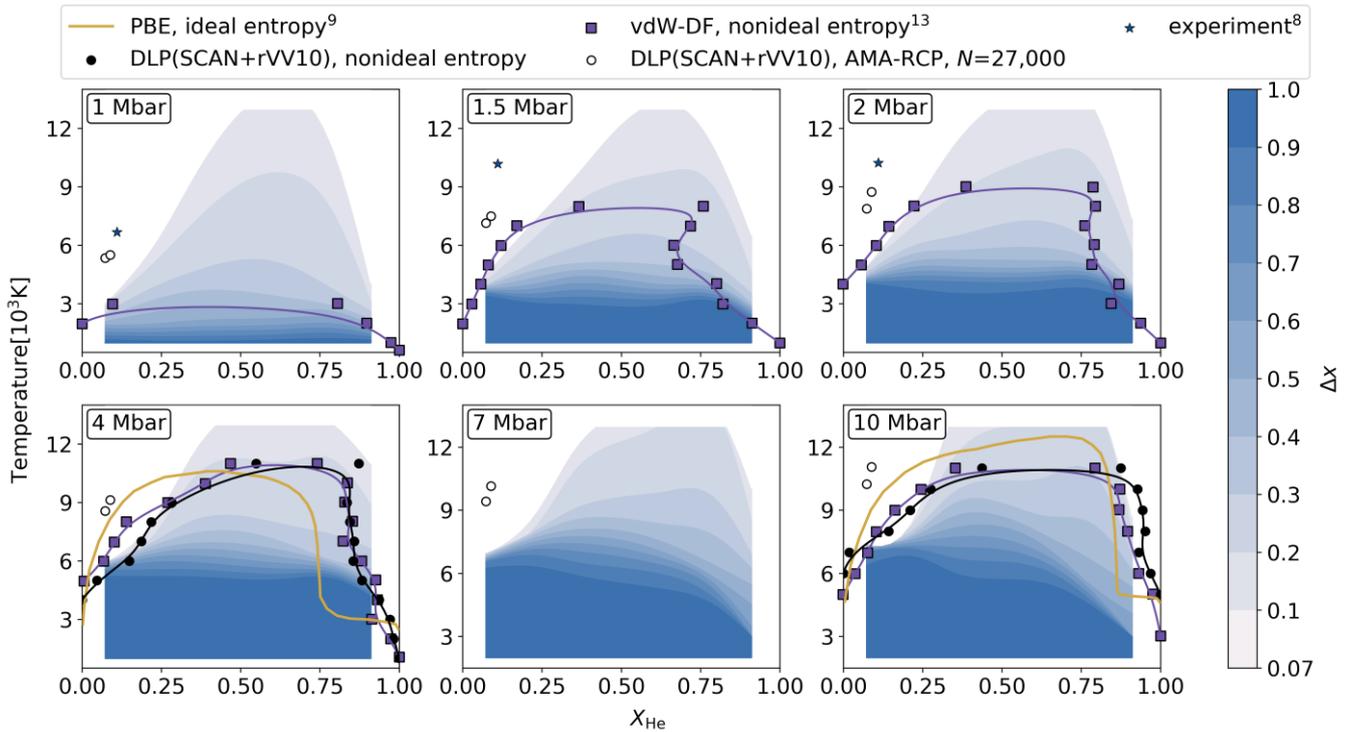

**Fig. 2. $\Delta x$ contour line diagram of H-He. a, b, c, d, e,** and **f** correspond to pressures of 1, 1.5, 2, 4, 7, and 10 Mbar. The color-coded areas represent different $\Delta x$ domains obtained using the AMA-RCP method on 9,604 atoms. The cited works and black dots with lines are all miscibility boundaries determined by the negative curvature of $\Delta G$. The black and purple lines of are the fifth-order B-spline fits to the points. Black hollow dots are points on miscibility boundaries determined by AMA-RCP. Blue

stars are from experiment.

In Fig. 2, we present the $\Delta x$ contour line diagram across various compositions of H-He mixture calculated at different pressures. The measure of immiscibility, $\Delta x$, ranges from 0.07 to 1. It is worth mentioning that $\Delta x =0.07$ remains in a separate state, where the color part only represents the degree of separation, not the separation boundary. At 4 and 10 Mbar, the miscibility lines (black dots) are calculated using thermodynamic integration on 64-atom systems to compare with the previous studies. Our results show a good consistence with the results obtained using vdW-DF and non-ideal entropy[13], with the miscibility lines shifting slightly toward higher temperature at the He-rich zone. This trend is more apparent in the comparison to results obtained using the PBE and ideal entropy[9]. These shifts are the consequence of non-ideal effects in entropy, which lowers the demixing temperatures at the He-poor zone[13] and captures the weak proton pairs at the He-rich zone leading to a higher demixing temperature[12]. Furthermore, the $\Delta x$ profiles reveal a rather complex landscape of H-He immiscibility driven by multiple factors, including temperature, pressure, and atomic compositions. The boundary of each $\Delta x$ domain is shown as a function of $X_{He}$, which is smoothed using a 5$^{th}$ order polynomial regression. These boundaries show a general trend of humping up in the middle. This suggests that the increase of temperature at a fixed pressure will lead to H-He mixing but the temperature required in the intermediate $X_{He}$ region is higher than that for the two ends. Moreover, the temperature-induced mixing is not symmetric with respect to $X_{He}$ – it has different gradients in He-poor and He-rich regions. As the pressure increases, the miscibility gap widens, and the immiscibility intensifies. However, the changes gradually slow down after the system has reached 7 Mbar; the miscibility diagram obtained at 10 Mbar sees much smaller changes from the former. It is observed that the immiscibility of hydrogen and helium is highly sensitive to helium abundance in planetary conditions. Once demixing is initiated even by a very weak miscibility gap, the accumulation of helium will accelerate the process. When the slow sedimentation of helium reaches a sufficient concentration, it might trigger stronger separation.

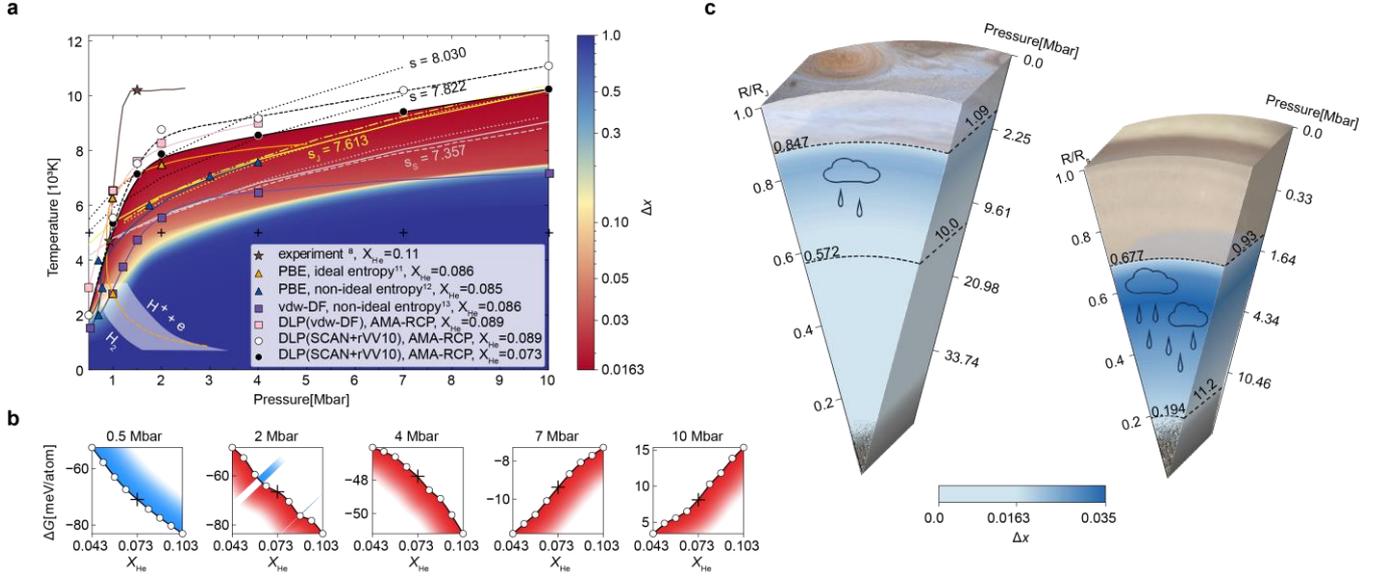

**Fig. 3. Miscibility diagram, ΔG, and implications for Jupiter and Saturn. a,** Miscibility diagram at various helium abundances. The solid lines of matching colors are the cubic B-spline fits to the calculation data points. The area under the immiscibility line of $X_{He} = 0.073$ is color-coded by the value of $\Delta x$, where 0.0163 is considered ahe threshold of immiscibility (see Methods). The color bar indicates the data values on a logarithmic scale. The two dotted black lines are calculated adiabats corresponding to their respective labeled entropy values. Solid yellow and pink lines are calculated adiabats of Jupiter and Saturn, presented alongside with previous results, dash-dot[23] and dotted[24] yellow lines for Jupiter, and dotted[24] and dash[25] pink lines for Saturn. According to previous studies[26–31], the white region under 3,000 K represents probable phase of hydrogen metallization. **b,** $\Delta G$ around the helium abundance of Jupiter and Saturn from Jarzynski equality. The five graphs correspond to the $\Delta G$ in a zoom-in abundance region centered on 0.073 calculated at five *P-T* points. Blue color represents stable regions with positive curvature and red color represents unstable and metastable regions with negative curvature. **c,** Implications for Jupiter and Saturn from **a**. The intervals enclosed by the dashed lines are H-He demixing layer of (left) Jupiter and (right) Saturn. The depth of blue represents the value of $\Delta x$ which measures the intensity of H-He phase separation.

Fig. 3a shows the miscibility diagram with current and previously published results. The calculated immiscibility line at the protosolar helium abundance ($X_{He}$ = 0.089) is shifted to higher temperatures compared with previous theoretical results[11–13], and it has the closest match to the experimental estimate[8]. Moreover, different helium abundances used in previous calculations and experiment should be noted. The four non-ideal adiabats are calculated using thermodynamic integration with the entropy of $s_J$ = 7.613 $k_B$ atom$^{-1}$ for Jupiter[23,24] and $s_S$ = 7.357 $k_B$ atom$^{-1}$ for Saturn[24,32], and two other entropy values, 7.822 $k_B$ atom$^{-1}$ and 8.030 $k_B$ atom$^{-1}$. The calculated demixing temperature (black and white dots) increases sharply upon increasing the pressure from 0.5 Mbar to 2 Mbar, coinciding with hydrogen metallization, and then flats out to a slow slope. This trend is the same as shown by DFT calculations[12,13] and experiment[8]. According to the miscibility diagram and adiabats, Jupiter would enter the miscibility gap at a pressure of 1.09 (±0.03) Mbar and a temperature of 5,440 (±50) K. For Saturn, the values are 0.93 (±0.03) Mbar and 4,720 (±50) K. The demixing region in Jupiter ceases at about 10,000 K when pressure is 10 Mbar, the highest pressure considered in this study, where Saturn is still in miscibility gap.

The miscibility gap inferred from structural analysis spans a very wide thermodynamic range, yet the intensity of immiscibility is relatively weak. This implies that the sedimentation of helium could be a protracted and subtle process. In Fig. 3a, the immiscibility line calculated for $X_{He}$ = 0.089 corresponds to the protosolar helium abundance. When the pressure exceeds 1 Mbar, this immiscibility line is notably higher than that of $X_{He}$ = 0.073, which is coherent to the positive slope of $\Delta x$ in He-poor region (Fig. 2). Based on the isentropes of various surface temperatures at protosolar helium abundance[14], the calculated immiscibility line for $X_{He}$ = 0.089 suggests that the onset of H-He phase separation in Saturn is about 0.746 Gyr (See Fig. S15). Due to the early onset and gradual accumulation, Saturn might experience a significant phase separation after sufficient quantitative change, whereas Jupiter would require a much longer time. The miscibility diagram calculated at protosolar helium abundance is provided in Fig. S15. To assess whether different functionals fundamentally influence the conclusions, we also trained a DLP based on vdW-DF and conducted identical simulations under protosolar helium abundance, as indicated by the pink squares in Fig. 3a. It can be observed that changing the functional still yields similar conclusions. The miscibility diagram calculated at protosolar helium abundance by vdW-DF based DLP is provided in Fig. S16.

To get an energy perspective, we calculated the $\Delta G$ using the systems of the same size (27,000 atoms) in a zoom-in helium abundance region at five pressures (0.5, 2, 4, 7, and 10 Mbar) and temperature of 5,000 K (Fig. 3b). The zoom-in region includes planetary helium abundance ($X_{He}$ = 0.073, plus symbol) and 8 adjacent points (circles). The calculation employs the Jarzynski equality[33] to a virtual integrable system[34], which enables the calculation of the Gibbs free energy in nonequilibrium states prior to demixing in large systems (See Methods for calculation of $\Delta G$). At 0.5 Mbar, all points on the $\Delta G$ curve have positive curvature (second derivative), suggesting that the state is stable without demixing. At 2 Mbar, small negative curvatures appear in $\Delta G$, signaling the occurrence of demixing. The demixing region is determined by common tangent construction, which is separated into unstable and metastable regions (red area in Fig. 3b). At 4 Mbar, although the $\Delta G$ curve still exhibits a downward trend, the negative curvature becomes quite evident – the phase separation is prevalent. At 7 Mbar, the $\Delta G$ curve starts to rise, and due to the constraint $\Delta G(0) = \Delta G(1) = 0$ (Eq. 2), there will be at least two minima where the system is stable. At 10 Mbar, $\Delta G$ curve rises to positive values, indicating the presence of strong separation in the system. These results correlate the change in curvature of $\Delta G$ to the demixing. Phase separation initially emerges in localized regions of negative curvature, and progressively proceeds as the line shape evolves.

Combining the calculated miscibility diagram with a planet model[3], and assume the interior of the planet is adiabatic, we derived the nowadays internal structures of Jupiter and Saturn (Fig. 3c). In Jupiter, the liquid layer of hydrogen, where helium droplets condensate and rain down, is estimated to be in between 0.572 and 0.847 of the radius. The calculated starting point for this layer is very close to the experimental estimate (0.84)[8] while it has a deeper ending point than the composition-corrected experimental estimate (0.68)[8]. This discrepancy is due to the different behaviors of demixing temperature predicted by theory and experiment (Fig. 3a). While the demixing temperature remains constant above 2 Mbar in experiment[8], it has a slow but continuous increase in theory. In Saturn, the separation layer is estimated to be in between 0.194 and 0.677 of the radius, indicates that nearly half of Saturn's radial distance would have helium rain. While the outer molecular envelope may still be well approximated as adiabatic, the deep interior is expected to depart from an adiabat due to helium rain and associated double-diffusive convection effects. Modeling these non-adiabatic regions is crucial for accurately describing the H-He demixing in Jupiter and Saturn. Incorporating these complexities into the model can lead to temperature shifts in the predicted hydrogen-helium separation region[5,35]. This work offers more precise parametric conditions for intricate planetary modeling. The *ab initio* H-He phase diagram delineates the *P-T* conditions at which helium becomes immiscible with hydrogen. It is crucial for determining the equilibrium helium abundance, influencing the depletion of the outer molecular envelope as rainout proceeds. The helium gradient, driven by the increasing helium concentration inward influences the profile of the double-diffusive convection region. The energy released from helium differentiation scales with the amount of helium that rains out, which is controlled by the phase diagram demarcating immiscible versus miscible regions. Through our large-scale calculations and structural analysis, we can also deduce the density ratios from the degree of demixing. This parameter can be used to parameterize the efficiency of heat transport by double-diffusive convection in the region where helium is raining out. We hope that our presented data can be incorporated into future planetary physics research to enhance the evolutionary modeling of planets.

## Discussion

In this work, we present direct observations suggesting the H-He immiscibility from calculations of *ab initio* quality. The structural analysis on machine learning accelerated large scale MD simulations establishes a renewed miscibility diagram of hydrogen and helium under the planetary conditions of Jupiter and Saturn. The immiscibility of hydrogen and helium is shown to be strongly dependent on temperature, pressure, and atomic composition of the mixtures. The miscibility gap is recalculated using the $\Delta G$ with the systems of the same size, which reinforces the structural analysis. Based on the results, we propose a new hypothesis for the mechanism of H-He separation in Jupiter and Saturn – the subtle increase in helium abundance through gradual helium accumulation in the early stage of planet formation potentially initiates the H-He phase separation. This process is similar to the formation of rain droplets in gathered clouds. The slightly larger immiscibility and lower adiabatic temperature within the interior of Saturn make it more likely to undergo this process. In contrast, in Jupiter the speed of this pre-sedimentation is lower, along with its higher adiabatic temperature decreases the likelihood of further separation.

## Methods

**DLP potential training**

DLP is constructed based on SCAN+rVV10. This scheme is chosen for its 'best-of-both-worlds' feature – SCAN captures the short and intermediate-range interactions and rVV10 describes the long-range vdW interaction[36]. The combination of SCAN+rVV10 has been shown to have accuracy better than 1 kcal/mol for benchmarking the MP2+ΔCCSD(T) results and be able to reproduce subtle features of the potential energy surface[37]. In addressing gigapascal pressures, SCAN+rVV10 has been shown to improve the agreement with experimental data compared with PBE and SCAN on liquid-liquid phase transition in nitrogen-oxygen mixtures, thanks to the revised noncovalent interactions of short- to long-range van der Waals interaction[38]. Similar better match appears in the works of Tantardini[39] and Anh[40]. Regarding metallic materials, SCAN+rVV10 has been demonstrated to align well with experimental results[41], outperforming PBE, PBEsol, and SCAN. Notably, the addition of rVV10 to SCAN yields a highly accurate method for diversely bonded systems.

We use the projector augmented wave[42] (PAW) approach for electron-ion interaction, specifically the hard PAW pseudopotentials for hydrogen (H_h, 06Feb2004), and helium (two valence electrons, He

05Jan2001) as provided with VASP. The DFT convergence tests are provided in Supplementary Information Note 1.

We use Deep Potential Generator (DP-GEN)[43] to cover temperatures from 1,000 K to 13,000 K and pressures from 0.5 Mbar to 10 Mbar. 28 helium proportions evenly distributed from 0 to 1 were used as training set. A total of 106,817 configurations in the training set, and trained for 12,000,000 steps.

**Explaination of re-weighting procedure**

In Fig. 1, after assigning a $P_{\text{cond}}$ value to each atom, we obtained atoms in their respective enriched regions (iv, $P_{\text{cond}} > 0.5$) and atoms in another element-enriched regions (iv, $P_{\text{cond}} < 0.5$). This way, we preliminary classified all atoms; however, atoms belonging to the same category may not have equal contributions to their respective regions. For instance, in the 6, 000K case in iii, a helium atom with $P_{\text{cond}} = 0.9$ contributes differently than a helium atom with $P_{\text{cond}} = 0.55$ (although both values are larger than 0.5). While the latter also appears in the helium-rich region, it behaves more like an interface atom. Therefore, it is necessary to further consider their weights when counting. To address this issue, we re-weighted atoms in the self-enrichment zone by the corresponding frequency of $P_{\text{cond}}$ values. And because of the causal links between two types of atoms within the same zone, the weights of atoms in the alternative element-enriched area should also be defined by the relevant $P_{\text{cond}}$ values frequency of the enriched type. As a result, the weights of all atoms exhibit central symmetry, as depicted by the grey shadows in iii. The effect of the re-weight procedure is visibly noticeable in iv, the weights of atoms at the interface decrease, while the weights of the genuine "shapers" are intentionally retained.

In Fig. S13, the values of $\Delta x$ corresponding to helium abundances ranging from pure hydrogen to pure helium under the condition of complete mixing are provided, which should be 0 under the assumption of perfect ideal conditions for infinite systems. However, in practical applications involving finite systems, where the dividing lines of $X_{\text{He}}$ and $1 - X_{\text{He}}$ are not sharp but have a Gaussian width, this introduces a certain level of uncertainty to the values of $x_1$ and $x_2$. As a result, achieving a fully mixed $\Delta x$ value of 0 is not feasible. And it can be observed that the order closer to the ideal scenario is (27,000 atoms, re-weight) > (27,000 atoms) > (9,604 atoms, re-weight) > (9,604 atoms). To enhance accuracy, it is necessary to employ as many atoms as possible and subsequently reweight the results of conditional probabilities.

**Immiscibility definition**

Through re-weighted conditional probability classification, we gain the $\Delta x = x_2 - x_1$ as a measure of immiscibility. Under the assumption of perfect ideal conditions for infinite systems, $\Delta x = 0$ means fully mixed and $\Delta x = 1$ means fully separated. However, in practical applications involving finite systems, when the atom number of the system is once fixed, the fully mixed $\Delta x$ value is typically correlated with $X_{He}$ of the system. Thus, the individualized immiscibility limits are considered for different $X_{He}$ in this paper. The distribution of $\Delta x$ corresponds to normal distribution through the Kolmogorov-Smirnov test[44], as shown in Fig. S13. 20,000 frames for each concerned composition were built randomly to better construct normal distributions of the mixed system. At the same time, trajectories from MD simulation were randomly shuffled to achieve a mixed system. From the analysis of these coordinates, the mean value of $\Delta x$ shows to be at largest at $X_{He} = 0.5$ and accelerates approaching 0 with the system's composition tends to be extreme. To explain this, we have to first assume a non-ideal mixed system with helium fraction $X_{He}$ which is misjudged separated because of random fluctuation. The fake $x_1$ and $x_1$ will be very close and are symmetric concerning the $X_{He}$:

$$x_1 + x_2 = 2X_{He}, \qquad (3)$$

the definition of $\Delta x$ is:

$$\Delta x_{mix} = x_2 - x_1, \qquad (4)$$

with Eq. 1 in the main text, and introduce fluctuation variable $\delta$ for non-ideal mixed system, it can be derived that:

$$\Delta x_{mix} = 2x_2 - 2X_{He} = \frac{2}{1 + \frac{N(H_{He\text{-rich}})}{N(He_{He\text{-rich}})}} - 2X_{He}$$

$$= \begin{cases} \dfrac{2}{1+\frac{1-X_{He}}{X_{He}}} - 2X_{He} = 0, & \text{ideal mix} \\ \dfrac{2}{1+\frac{1-X_{He}}{X_{He}}\delta} - 2X_{He}, & \text{non-ideal mix} \end{cases}. \qquad (5)$$

In ideal mixed system, the He-rich region completely overlaps with the He-poor region, thus $\frac{N(H_{He\text{-rich}})}{N(He_{He\text{-rich}})}$ is equal to $\frac{1-X_{He}}{X_{He}}$. For non-ideal mixed systems, a fluctuation variable $\delta$ has to be added to thought. The dashed function fit lines are from Eq. 5 with proper $\delta$ to fit random and shuffled trajectories. The relation between $\overline{\Delta x_{mix}}$ and $X_{He}$ of mixed system indicates different demixing definitions for systems with different compositions.

The inset of Fig. S13 shows the distribution of $\Delta x_{mix}$ of 27,000 atoms. Because $\Delta x_{mix}$ of mixed system abides by normal distribution, the three-sigma rule[45] is used to determine the mixed interval. $\mu$ is the mean of the distribution, and $\sigma$ is its standard deviation. $\mu + 3\sigma$ values are marked in the figure as the right boundaries of intervals, in which 99.865% of $\Delta x_{mix}$ will lie in. Now, with a $\Delta x$ calculated from a single frame or a $\overline{\Delta x}$ from a trajectory containing many frames, we can compare it with $\mu + 3\sigma$ value of $\Delta x_{mix}$. If the value falls outside of the $3\sigma$ intervals, indicating that the probability of the system being mixed is only 0.135%. For a $\Delta x$ from a single frame, it risks with these small probabilities, but for a $\overline{\Delta x}$ from a long trajectory of a wide time scale containing many frames, this risk can be further reduced. Thus, the $\mu + 3\sigma$ values of $\Delta x_{mix}$ for every $X_{He}$ are used as the immiscibility threshold. The inset of Fig. S13 presents an application example of immiscibility definition at helium abundance of Jupiter and Saturn along 2 Mbar. With the decrease of the temperature, the distributions of $\Delta x$ gradually move away from the top mixed one. When the mean value of the distribution is greater than $\mu + 3\sigma$, the system is determined to be in the miscibility gap, and so is Jupiter and Saturn's case at this thermodynamic condition.

The $\mu + 3\sigma$ values for systems with 9,604 and 27,000 atoms at certain helium abundances are shown in the Table. S1. For binary systems, the results are symmetric around a helium abundance of 0.5, so we only list the values for $X_{He} < 0.5$.

**MD simulations details.**

To analyze through structures of trajectories, two sets of MD simulations were performed. The first set of calculations was carried out on 9,604-atom simulation boxes at compositions of $X_{He}$ = 0.2, 0.357, 0.5, 0.643, 0.8, and 0.9. The second set of calculations was performed on 27,000-atom simulation boxes with helium abundances of 0.073 (representing the current helium abundance in the atmospheres of Jupiter and Saturn) and 0.089 (representing protosolar helium abundance, which is the theoretical initial helium abundance of Jupiter and Saturn). For each $X_{He}$, calculations were performed at pressures of 0.5, 1, 1.5, 2, 4, 7, and 10 Mbar, and necessary temperatures ranging from 1,000 to 13,000 K. A timestep of 0.2 fs was used and 100 ps MD simulation was performed to guarantee a long enough trajectory for full sample after thermodynamic equilibrium was reached.

**$\Delta G$ calculation of large systems.**

When calculating the $\Delta G$, it is crucial to ensure that the term $G(X_{He})$ in Eq. 2 refers to a fully mixed system, as it cannot represent the Gibbs free energy of a specific helium abundance $X_{He}$ once the system undergoes demixing. Unlike previous work using small enough systems to guarantee the fully mixed state in simulation, we apply the Jarzynski equality to a virtual integrable system[34] to calculate the Gibbs free energy difference of volume change:

$$e^{-\beta(G_B - G_A)} = (V_B/V_A)^N \langle e^{\beta(U_A - U_B)} \rangle_{A,\mathbf{r}}. \qquad (6)$$

Here $\beta = 1/k_B T$ is the inverse temperature, $U_A$ and $U_B$ are the interaction energies of the initial state A and final state B, and the corresponding Gibbs free energies are represented by $G_A$ and $G_B$. $\mathbf{r} = (x, y, z)$ is the coordinates of all particles at initial state A. Since the transition from state A to state B only involves a change in volume while keeping the relative positions of all atoms unchanged, the miscibility of state B remains the same as state A. Therefore, we only need to locate a completely mixed point on the miscibility diagram to be the initial state A. The calculation process for determining $\Delta G$ in large systems can be roughly divided into two steps: 1) calculating the absolute free energy of completely mixed state A using thermodynamic integration, and 2) repeatedly sampling the process from state A to state B using Eq.6.

After considering structure analysis results of this work, as well as multiple previous calculations and experiments, we have selected the temperature of 5,000 K and pressure of 0.5 Mbar as the condition for the completely mixed state A, which can be observed from Fig. 2. Moreover, at a temperature of 5,000 K, there will be various degrees of phase separation occurring with changes in pressure. This provides an opportunity for more comprehensive validation.

To compare systems with different particle compositions, it is necessary to calculate the absolute Gibbs free energy. We calculate the absolute Gibbs free energy of 11 components (including pure hydrogen, pure helium, and components near the abundance of planetary helium) at the initial state of 0.5 Mbar and 5,000K through thermodynamic integration (TI). As a result, we obtain the $\Delta G$ curve representing the completely mixed initial state. Then we apply the Jarzynski equality to a virtual integrable system[34] to calculate Gibbs free energy difference of volume change and repeat this sampling process to calculate the ensemble average on the right-hand side of Eq. 6. Subsequently, several sets of $\Delta G$ calculations were performed near the planetary helium abundance along the 5,000 K isotherm in the *P-T* miscibility diagram.

By taking the logarithm of both sides of Eq. 6 and considering values per atom, we obtain:

$$\frac{-(G_B - G_A)}{N} \approx \frac{V_B}{\beta V_A} + \frac{\langle U_A - U_B \rangle_{A,\mathbf{r}}}{N}. \qquad (7)$$

Thus, the error of $\Delta G$ mainly comes from two parts: $\frac{V_B}{\beta V_A}$ and $\frac{\langle U_A - U_B \rangle_{A,\mathbf{r}}}{N}$. Generally, our calculations are based on a system with 27,000 atoms, however, when relaxing to obtain the $V_B$, we employ larger system sizes. 729,000 atoms systems are used for 2 Mbar while 216,000 atoms systems are used for 4 Mbar, 7 Mbar, and 10 Mbar. These larger sizes are 27 times and 8 times the target size, respectively (The smaller the pressure, the larger the volume with fluctuation, and the greater the number of atoms required to maintain consistent accuracy). In this way, $\frac{\sigma(\overline{V_B})}{\beta V_A}$ is of the order $10^{-7}$ eV. As for repeated sampling of $\frac{\langle U_A - U_B \rangle_{A,\mathbf{r}}}{N}$, we have performed convergency test on sample size to ensure the accuracy. See Fig. S14.

**$\Delta G$ calculation by thermodynamic integration.**

Thermodynamic integration calculates differences in free energies to get the target state (described by potential $U_1$) free energy integrated from a reference state (described by potential $U_0$). Then define a transition potential with two switching functions $f(\lambda), g(\lambda)$:

$$U = f(\lambda)U_0 + g(\lambda)U_1, \qquad (8)$$

$f(\lambda)$ and $g(\lambda)$ satisfy $f(0)=1, f(1)=0, g(0)=0$ and $g(0)=1$. With Eq. 8, the free energy $F$ is a function of $N, V, T, \lambda$:

$$F(\lambda) = -k_B T \ln Q(N,V,T,\lambda). \qquad (9)$$

The difference between the target state and reference state is:

$$F_1 - F_0 = \int_0^1 d\lambda \frac{\partial F(\lambda)}{\partial \lambda}, \qquad (10)$$

And it can be derived that $\frac{\partial F(\lambda)}{\partial \lambda} = \left\langle \frac{\partial U}{\partial \lambda} \right\rangle_\lambda$ through:

$$\frac{\partial F(\lambda)}{\partial \lambda} = -\frac{k_B T}{Q}\frac{\partial Q(\lambda)}{\partial \lambda} = -\frac{k_B T}{Z}\frac{\partial Z(\lambda)}{\partial \lambda} = -\frac{k_B T}{Z}\frac{\partial}{\partial \lambda}\int d^N\mathbf{r}\, e^{-\beta U(\mathbf{r}_1,\ldots,\mathbf{r}_N,\lambda)}$$

$$= -\frac{k_B T}{Z}\int d^N\mathbf{r}\left(-\beta\frac{\partial U}{\partial \lambda}\right)e^{-\beta U(\mathbf{r}_1,\ldots,\mathbf{r}_N,\lambda)} = \frac{\int d^N\mathbf{r}(\frac{\partial U}{\partial \lambda})e^{-\beta U(\mathbf{r}_1,\ldots,\mathbf{r}_N,\lambda)}}{\int d^N\mathbf{r}\, e^{-\beta U(\mathbf{r}_1,\ldots,\mathbf{r}_N,\lambda)}} = \left\langle \frac{\partial U}{\partial \lambda} \right\rangle_\lambda. \qquad (11)$$

Replace the corresponding term in Eq. 10:

$$F_1 - F_0 = \int_0^1 d\lambda \left\langle \frac{\partial U}{\partial \lambda} \right\rangle_\lambda. \qquad (12)$$

From Eq. 12, the difference of free energies can be obtained by integrating transition potential $U$ defined by Eq. 8. In this work, the switching functions were defined as $f(\lambda) = 1 - \lambda$ and $g(\lambda) = \lambda$. Ideal gas was taken as the reference state.

Ideal gas is chosen as the reference state, from which we integrate 47 coupling constants[10] to the target state with DLP potential. After a relaxation of at least 50 ps of a certain composition to thermodynamic equilibrium, all MD simulations for integration have been run for 2 ps with a timestep of 0.2 fs. 7 temperatures and 64 compositions are considered for the two concerned pressures. We also calculate two $\Delta G$ lines with vdW-DF and SCAN+rVV10 using VASP[46] code at 10 Mbar and 5000 K to benchmark the comparison.

With free energy $F$ of a certain fully mixed system with helium fraction $X_{\text{He}}$, the delta Gibbs free energy can be calculated by Eq. 2, where $G(X_{\text{He}}) = F(X_{\text{He}}) + PV$. Then, by applying a common double tangent construction to $\Delta G$ we can determine $x_1$ and $x_2$. According to Fig. S17, the negative curvature of $\Delta G$ gradually disappear with the increase of temperature.

## Data availability

The relevant data for this research is accessible at:

lurr. (2024). changxiaoju/H-He-Immiscibility: v1.0.0 (v1.0.0). Zenodo.

https://doi.org/10.5281/zenodo.13120470.

## Code availability

LAMMPS and DeepMD are free and open source codes available at https:// lammps.sandia.gov and http://www.deepmd.org, respectively. VASP is a commercial code available from https://www.vasp.at. Detailed instructions for obtaining and using these codes can be found on their respective websites. The AMA-RCP code is provided in https://doi.org/10.5281/zenodo.13120470.

**Acknowledgments** This work was supported by the National Key R&D Program of China under Grant No. 2017YFA0403200, the National Natural Science Foundation of China under Grant Nos., No. 12047561, and No. 12104507, the NSAF under Grant No. U1830206, the Science and Technology Innovation Program of Hunan Province under Grant No. 2021RC4026. We acknowledge Jianmin Yuan, Yexin Feng, and Rei Cao for their helpful discussions.

**Author contributions** J.D. designed the project. X.C., B.C. Y.Y. Y.M. and J.D. suggested the specific scientific problem and the general idea on methodology, X.C. and J.D. performed the MD simulations and analyzed data, K.C. and H.W. contributed to the AMA-RCP. X.C., B.C., Q.Z., H.W., K.C., Q.T., X.Y.,

D.K., S.Z., F.G., Y.H., Z.Z, Y.Y, Y.M, and J. D. interpreted the results, X.C., Y.Y., and J.D. wrote the paper, and Q.Z., H.W., K.C., Q.T., X.Y. D.K., Y.H., Z.Z, Y.M, X.Y., S.Z., and D.K. edited the manuscript before submission.

**Competing interests** The authors declare no competing interests.

**Additional information**

**Supplementary information** is available for this paper at www.